\documentclass[journal]{IEEEtran}
\usepackage{algorithm}
\usepackage{eurosym}
\usepackage{algpseudocode}
\usepackage{amsfonts}
\usepackage{amsmath}
\usepackage{nicefrac}
\usepackage{array}
\usepackage{booktabs}
\usepackage{stfloats}
\usepackage{graphicx}
\usepackage[hidelinks]{hyperref}
\graphicspath{ {D:/Paper2/Drafts/Figures/} }
\usepackage{nomencl}
\usepackage{etoolbox}
\renewcommand\nomgroup[1]{%
  \item[\bfseries
  \ifstrequal{#1}{R}{Robust UC}{%
  \ifstrequal{#1}{F}{SFR model}{%
  \ifstrequal{#1}{A}{Acronyms}{%
  \ifstrequal{#1}{L}{LR model}{}}}}%
]}
\makenomenclature
\usepackage{xcolor,colortbl,color,soul}
\definecolor{myRed}{RGB}{243, 129, 129}
\definecolor{myYellow}{RGB}{252, 227, 138}
\definecolor{myGreen}{RGB}{234, 255, 208}
\definecolor{myTeal}{RGB}{149, 225, 211}
\usepackage{tabularx}
\newcolumntype{g}{X}
\newcolumntype{n}{>{\hsize=.8\hsize}X}
\newcolumntype{s}{>{\hsize=.48\hsize}X}
\usepackage{multirow}
\usepackage[framemethod=tikz]{mdframed}
\usepackage{tikz}
\usepackage{booktabs}
\usetikzlibrary{calc}
\newcommand{\tikzmark}[1]{\tikz[overlay,remember picture] \node (#1) {};}
\newcommand{\DrawBox}[3][]{%
    \tikz[overlay,remember picture]{
    \draw[black,#1]
      ($(#2)+(-0.5em,1.3ex)$) rectangle
      ($(#3)+(0.5em,-0.25ex)$);}
}

\usepackage[normalem]{ulem} % to take out sections: \sout
\usepackage[draft,nompar]{commenting}  % of [draft], [final
\declareauthor{Lukas}{Lukas}{red} 
\authorcommand{Lukas}{comment} % \Lukas{}

\begin{document}
\title{Robust Frequency Constrained UC Using Data Driven Logistic Regression for Island Power Systems}

\author{Mohammad~Rajabdorri,
        Enrique~Lobato,
        Lukas~Sigrist,~\IEEEmembership{Member,~IEEE}
\thanks{This study is funded by European Regional Development Fund (ERDF), Ministerio de Ciencia e Innovación - Agencia Estatal de Investigación, Project RTI2018-100965-A-I00.\\
The authors are with IIT School of Engineering ICAI Universidad Pontificia Comillas, Madrid, Spain. (e-mails: mrajabdorri@comillas.edu, enrique@comillas.edu, lsigrist@comillas.edu)}
}

\markboth{Second submitted draft}{-}

\maketitle

\begin{abstract}
In the current practice of short-term power scheduling, online power reserves are used to address generation mismatches and contingencies. Neither online inertia nor the speed of the committed units is considered in the scheduling process. With the increasing injection of uncertain renewable energy sources, this practice is starting to fall short especially in island power systems, where the primary frequency response is already scarce, and any contingency leads to potentially poor frequency response. This paper introduces a data driven linear constraint to improve the post-fault frequency quality in island power systems. A coherent initial data-set is obtained by simulating the system frequency response of single outages. Then logistic regression is employed as a predictive analytic procedure to differentiate the acceptable and unacceptable incidents. To compare the conventional methods with the proposed approach and also to handle the uncertain nature of renewable energy generation, an adaptive robust unit commitment formulation is utilized. Results for the island power system of La Palma show that depending on the chosen cut-point on the logistic regression estimation the proposed method can improve the frequency response quality of the system while reducing the operation costs.
\end{abstract}

\begin{IEEEkeywords}
Frequency Constrained Unit Commitment, Logistic Regression, System Frequency Response Model, Robust Optimization.
\end{IEEEkeywords}

\IEEEpeerreviewmaketitle

\nomenclature[R]{\(suc(.)\)}{Start-up costs [\euro]}
\nomenclature[R]{\(gc\)}{Generation costs [\euro]}
\nomenclature[R]{\(x\)}{Commitment variable [$\in$\{0,1\}]}
\nomenclature[R]{\(y\)}{Start-up variable [$\in$\{0,1\}]}
\nomenclature[R]{\(z\)}{Shut-down variable [$\in$\{0,1\}]}
\nomenclature[R]{\(p\)}{Power generation variable [MW]}
\nomenclature[R]{\(t\)}{Index of time intervals}
\nomenclature[R]{\(\mathcal{T}\)}{Set of all time intervals}
\nomenclature[R]{\(i\)}{Index of generators}
\nomenclature[R]{\(\mathcal{I}\)}{Set of all generators}
\nomenclature[R]{\(tt\)}{Alias index for time intervals}
\nomenclature[R]{\(ii\)}{Alias index for generators}
\nomenclature[R]{\(UT\)}{Minimum up-time of generators [hours]}
\nomenclature[R]{\(DT\)}{Minimum down-time of generators [hours]}
\nomenclature[R]{\(\overline{\mathcal{P}_i}\)}{Maximum power output of generator $i$ [MW]}
\nomenclature[R]{\(\underline{\mathcal{P}_i}\)}{Minimum power output of generator $i$ [MW]}
\nomenclature[R]{\(\overline{\mathcal{R}_i}\)}{Maximum ramp-up of generator $i$ [MW]}
\nomenclature[R]{\(\underline{\mathcal{R}_i}\)}{Maximum ramp-down of generator $i$ [MW]}
\nomenclature[R]{\(r\)}{Online reserve power variable [MW]}
\nomenclature[R]{\(\mathcal{W}\)}{Set of Wind generation uncertainty}
\nomenclature[R]{\(wg\)}{Wind generation variable [MW]}
\nomenclature[R]{\(w\)}{Available forecasted wind power [MW]}
%\nomenclature[R]{\(C\)}{Linear cost points [\euro/MW]}
\nomenclature[R]{\(\alpha\)}{Dual variable of minimum power constraint}
\nomenclature[R]{\(\beta\)}{Dual variable of maximum power constraint}
\nomenclature[R]{\(\gamma\)}{Dual variable of down ramp constraint}
\nomenclature[R]{\(\delta\)}{Dual variable of up ramp constraint}
\nomenclature[R]{\(\eta\)}{Dual variable of power balance constraint}
\nomenclature[R]{\(\zeta\)}{Dual variable of maximum wind constraint}
\nomenclature[R]{\(\mu\)}{Dual variable of minimum reserve constraint}
\nomenclature[R]{\(\rho\)}{Dual variable of LR constraint}
\nomenclature[F]{\(\mathcal{H}\)}{Inertia [s]}
%\nomenclature[F]{\(\widetilde{\mathcal{H}}\)}{Normalized inertia [s]}
\nomenclature[F]{\(\Delta\Dot{w}_i\)}{frequency deviation [p.u.]}
\nomenclature[F]{\(\Delta d\)}{Load deviation [p.u.]}
\nomenclature[F]{\(\widetilde{\Delta p}\)}{Total mechanical power deviation [p.u.]}
\nomenclature[F]{\(\widetilde{\Delta d}\)}{Total load deviation [p.u.]}
\nomenclature[F]{\(\Delta p\)}{Mechanical power deviation [p.u.]}
\nomenclature[F]{\(a_.\)}{Poles of the second order system}
\nomenclature[F]{\(b_.\)}{Zeros of the second order system}
\nomenclature[F]{\(k\)}{Inverse of the droop [p.u.]}
\nomenclature[F]{\(n\)}{Index of contingency}
\nomenclature[F]{\(N\)}{Total number of contingencies}
\nomenclature[F]{\(\mathcal{K}\)}{Normalized gain of turbine-governor model}
\nomenclature[F]{\(\mathcal{M}\)}{Base rated power of units [MW]}
\nomenclature[F]{\(\mathcal{S}\)}{Base power of the system [MW]}
\nomenclature[L]{\(\upsilon\)}{Dependant variable}
\nomenclature[L]{\(\xi\)}{Independent variable}
\nomenclature[L]{\(\pi (.)\)}{Probability of .}
\nomenclature[L]{\(c\)}{Regression coefficient}
\nomenclature[L]{\(\psi\)}{Regression cut-point}
\nomenclature[L]{\(f^{nadir}\)}{The minimum value of frequency reached during the transient period}
\nomenclature[L]{\(f^{qss}\)}{Quasi steady-state frequency}
\nomenclature[A]{\(RoCoF\)}{Rate of Change of Frequency}
\nomenclature[A]{\(UC\)}{Unit Commitment}
\nomenclature[A]{\(SFR\)}{System Frequency response Model}
\nomenclature[A]{\(ED\)}{Economic Dispatch}
\nomenclature[A]{\(UFLS\)}{Under Frequency Load Shedding}
\nomenclature[A]{\(RUC\)}{Robust Unit Commitment}
\nomenclature[A]{\(LR\)}{Linear Regression}
\nomenclature[A]{\(RES\)}{Renewable Energy Sources}
\nomenclature[A]{\(OCT\)}{Optimal Classification Trees}
\printnomenclature

\section{Introduction}
\begin{figure}[t]
\centering
\includegraphics[width=\linewidth]{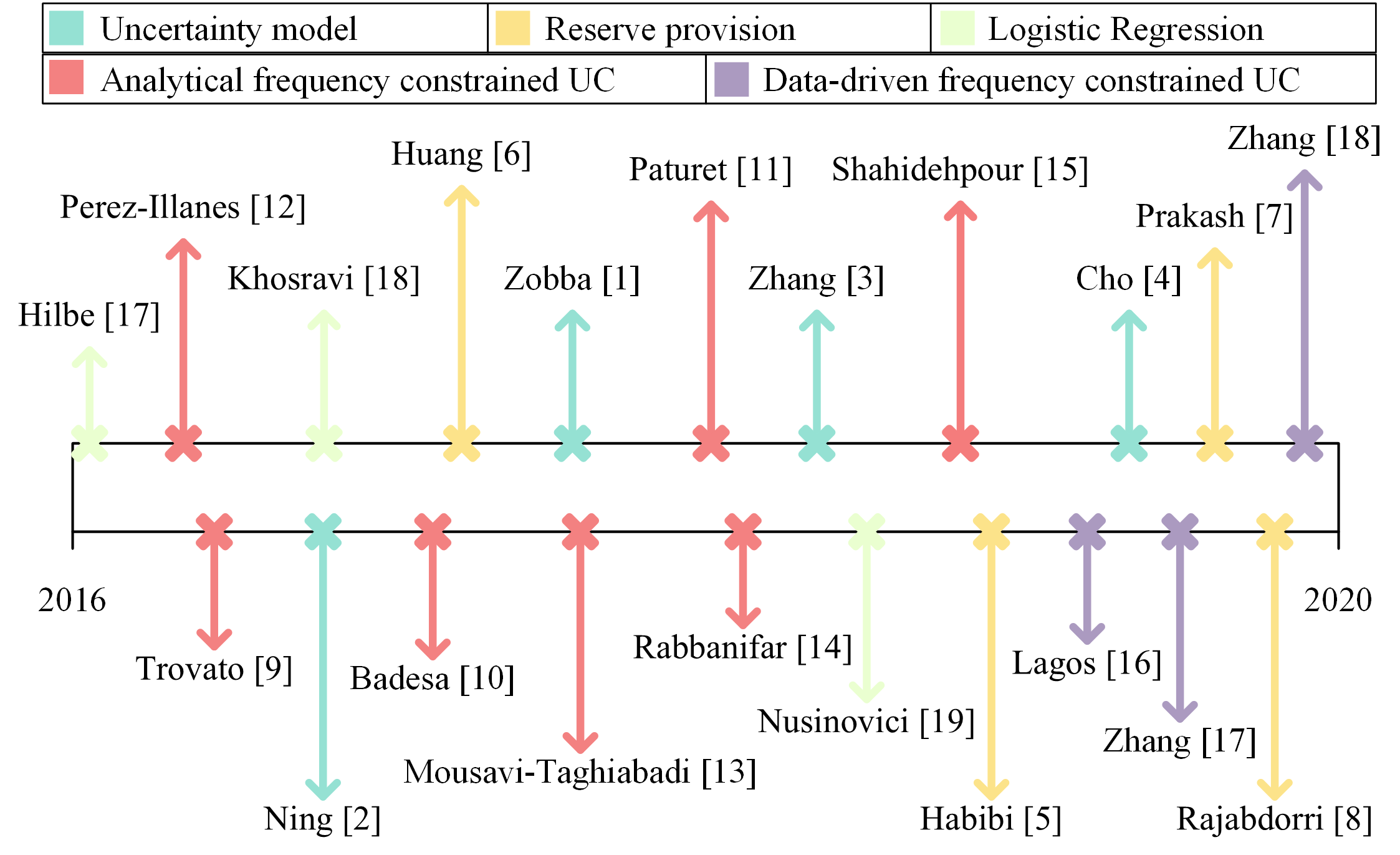}
\caption{Summary of references.}
\label{freferences}\end{figure}

\IEEEPARstart{V}{ariability} and uncertainty are becoming a bigger concern in power systems due to the ever-increasing penetration of RES as a source of power generation. Among power systems, island power systems suffer more as they inherently possess less inertia and primary frequency control capacity. Inertia scarcity in island power systems makes them more susceptible to power outages and fluctuations in uncertain renewable energy sources (RES). Traditionally, online reserve power provided by conventional units has been the main tool to tackle unforeseen sudden changes of power balance and to maintain the frequency within a tolerable range. The current reserve assignment is such that the N-1 criterion is covered and expected load and RES variations can be absorbed, but it ignores available inertia and system response speed. This practice is falling short as (1) the conventional units are less utilized by increasing the share of RES, (2) the amount of available reserve might not be enough depending on the changes in RES infeed, which is exposed to forecast errors, (3) the system is left with insufficient amount of responsive resources facing outages and forecast errors.\\
To address the volatile nature of RES and include the stochasticities in the scheduling process, usually stochastic and robust models are employed. Considering the pros and cons of different models, an adaptive robust UC is employed for the purpose of this paper. Some of the more recent usages and developments in the formulation can be found in \cite{zobaa2020uncertainties}, \cite{ning2019data}, \cite{zhang2020partition}, and \cite{cho2021three}. To ensure the provision of sufficient and fast reserves, different solutions are introduced in the literature (\cite{habibi2021assessment}, \cite{huang2019endogenous}, \cite{prakash2022frequency}, \cite{rajabdorri2021viability}). While new sources of reserve are being introduced, it's also essential to make sure that the quality of frequency transitions is guaranteed in the scheduling process, in case of any abrupt contingency.\\
Following the higher injection of RES to the grid, the larger frequency deviations are expected after any power mismatch. The amount of frequency control that is needed depends on system inertia, generation loss, and the speed of providing reserve. More attention is being paid to this issue. One obstacle is that frequency-related constraints, like frequency nadir, are highly non-linear, so it’s hard to implement them in the scheduling process, which is usually solved by mixed-integer linear programming methods. In \cite{trovato2018unit}, a linear formulation is introduced that equips the unit commitment problem with information about inertial response and the frequency response of the system and makes sure that in case of the largest outage, there is enough ancillary service to prevent under frequency load shedding (UFLS). To linearize frequency nadir constraint, first-order partial derivatives of its equation with respect to higher-order non-linear variables are calculated. Then the frequency nadir is presented by a set of piecewise linearized constraints. In \cite{badesa2019simultaneous}, different frequency services are optimized simultaneously with a stochastic unit commitment (SUC) approach, targeting low inertia systems that have high levels of RES penetration. The stochastic model uses scenario trees, generated by the quintile-based scenario generation method. To linearize frequency nadir, an inner approximation method is used for one side of the equation, and for the other side, a binary expansion is employed and linearized using the big-M technique. In \cite{paturet2020stochastic}, a stochastic unit commitment approach is introduced for low inertia systems, that includes frequency-related constraints. The problem considers both the probability of failure events and wind power uncertainty to compute scenario trees for the two-stage SUC problem. An alternative linearization approach is used to make sure the nadir threshold is not violated. Instead of piece-wise linearizing the whole equation, relevant variables including the nonlinear equation are confined within a plausible range that guarantees frequency drop after any contingency will be acceptable. Reference \cite{perez2016robust} has proposed a forecasting approach to model the uncertainties of RES to define upper and lower bounds and further implement them in a robust unit commitment (RUC). This study has assumed that frequency deviation is a linear function of the RoCoF, and has added it as a constraint to the RUC problem. In \cite{mousavi2020integration}, a reformulation linearization technique is employed to linearize the frequency nadir limit equation. To address the uncertainties of wind generation, an improved interval unit commitment is used. Results show that controlling the dynamic frequency during the scheduling process decreases the operation costs of the system while ensuring its frequency security. In \cite{rabbanifar2020frequency}, first, a frequency response model is developed that provides enough primary frequency response and system inertia in case of any outage. All frequency dynamic metrics, including the RoCoF and frequency nadir are obtained from this model, as analytic explicit functions of UC state variables and generation loss. These functions are then linearized based on a pseudo-Boolean theorem, so they can be implemented in linear frequency constrained UC problem. To find the optimal thermal unit commitment and virtual inertia placement, a two-stage chance-constrained stochastic optimization method is introduced in \cite{shahidehpour2021two}. Frequency nadir is first defined with a bi-linear equation and then it’s linearized with the help of the big-M approach.\\
In \cite{lagos2021data}, instead of extracting analytical formulas from swing equation, a data-driven multivariate optimal classification trees (OCT) technique is used to extract linear frequency constraints. A robust formulation is proposed to address the uncertainties of load and RES. A dynamic model is presented in \cite{Zhang2021} to generate the training data. The generated data is trained by deep neural network. Trained neural networks are formulated so they can be used in an MIL problem and the frequency nadir predictor is developed, to be used in UC problem. Then in \cite{Zhang2021a} DNN is trained by high-fidelity power simulation and reformulated as an MIL set of constraints to be used in UC. A summary of the reviewed FCUC related papers is provided in table \ref{refSummary}.\\
\begin{table}[t]
\caption{A Summary of Frequency constrained UC References}
\label{refSummary}
\centering
\begin{tabular}{>{\centering\arraybackslash}p{0.13\linewidth}|>{\centering\arraybackslash}p{0.17\linewidth}|>{\centering\arraybackslash}p{0.30\linewidth}|>{\centering\arraybackslash}p{0.2\linewidth}}
\toprule
$\#$/year &	Uncertainty model &	linearization technique &	Case study  \\
\hline
\rowcolor{myTeal}
\cite{trovato2018unit}/2018 & Deterministic &	First order partial derivatives	& Great Britain 2030 \\
\cite{badesa2019simultaneous}/2019 & Stochastic & Inner approximation and binary expansion & Great Britain 2030 \\
\rowcolor{myTeal}
\cite{paturet2020stochastic}/2020 & Stochastic & Extracting bounds on relevant variables & IEEE RTS-96 \\
\cite{perez2016robust}/2016 & Robust & Assuming nadir is a linear function of RoCoF & Northern Chile \\
\rowcolor{myTeal}
\cite{mousavi2020integration}/2020 & Improved interval & Reformulation linearization technique & IEEE 6-bus \\
\cite{rabbanifar2020frequency}/2020 & Deterministic & Pseudo-Boolean theorem & IEEE RTS-96 \\
\rowcolor{myTeal}
\cite{shahidehpour2021two}/2021 & Chance-constrained & Binary expansion & China 196-bus\\
\cite{lagos2021data}/2021 & Robust & Data-driven optimal classifier trees & Rhodes island and IEEE 118\\
\rowcolor{myTeal}
\cite{Zhang2021}/2021 & Deterministic & DNN trained by dynamic simulation & Modified 33-node system\\
\cite{Zhang2021a}/2021 & Deterministic & DNN trained by high-fidelity generated data & IEEE 39-bus system\\
\bottomrule
\end{tabular}
\end{table}
Analytical formulations for frequency metrics are usually based on simplifications of the non-linear behavior of power systems during large active power unbalance. To include the non-linear frequency metrics in linear UC, reviewed references are trying to employ a linearization technique. Eventually, the obtained linear lines are always a function of system dynamic constants, available inertia, and the amount of power imbalance. Although this serves the purpose of ensuring the quality of frequency response, it usually increases the size and complexity of the UC problem, in order to reach some level of accuracy. This paper employs Logistic regression (LR) as a dichotomous classification approach to classify the post-fault frequency drop as acceptable or unacceptable. LR is one of the most useful statistical procedures in healthcare analysis, medical statistics, credit rating, ecology, social statistics and econometrics, and etc. This procedure is important in predictive analytics, as it's able to categorize the outcome \cite{hilbe2016practical}. Considering the problem at hand and the purpose of this paper, this approach is promising. In \cite{khosravi2019expect}, a framework is proposed that removes irrelevant features with no effect on classification and concludes that a training data-set with missing values can still generate sufficient explanations of LR classifications. The standard LR model is compared with 5 different machine learning models to predict the risk of major chronic diseases in \cite{nusinovici2020logistic}. The results show that LR yields as good performance as other machine learning models. An advantage over other methods such as \cite{lagos2021data} or \cite{rabbanifar2020frequency} is that no additional decision variables are needed, maintaining model complexity. A summary of all discussed papers is shown in figure \ref{freferences}.\\
To the best of the authors' knowledge, logistic regression has not been used as an analytic tool in the UC problem and has never been employed to estimate the quality of frequency response in island power systems. Considering the presented background, this paper proposes a predictive analytic approach to enhance post-fault frequency quality in a robust UC model. The idea is to avoid dispatches that lead to poor frequency responses by scheduling only those generators whose outage would not violate acceptable frequency deviations, thus reducing the potential UFLS.\\
This paper proposes a novel data driven constraint, by analyzing a coherent data-set, using logistic regression procedure. To build an initial set of data to train the LR model, an adaptive robust UC formulation with reserve constraint is employed and solved for different levels of reserve requirement. The idea of using different levels of reserve is to simultaneously determine the level actually needed. The commitment variables of the robust UC solution for different levels of reserves are used to solve the economic dispatch (ED) problem for day-ahead stochastic scenarios. Every single outage of the obtained results is simulated by an SFR model, which makes the training data-set a proper representative of all acceptable and unacceptable frequency responses. From the training data-set a new constraint is derived using the logistic regression procedure, and then included in robust UC instead of conventional reserve constraint to ensure both frequency quality after outages, and feasibility of the result in case of any realization of the stochastic variable. Although the linearization happens in the training process, the new constraint does not add the number of constraints in UC problem, hence keeping the problem size intact. To compare the proposed approach with recent data-driven methods that are introduced in the literature, OCT is also used to train a linear constraint with. Both methods are compared in the results and their computational run-time and improvements in the frequency quality are highlighted.
Key contributions and merits compared to the current state of the art can be summarized as,
\begin{itemize}
    \item This paper introduces logistic regression as a tool to train output data of SFR model, and develops a new constraint to be used in UC problem instead of reserve constraint.
    \item Proposed formulation does not add any new binary, integer, or free variables to the UC problem and does not enlarge the number of UC constraints, conserving the size and complexity of the problem.
    \item The procedure of training the new constraint is very fast and can be done, using any computer algebra system.
\end{itemize}

The rest of the paper is organized as follows. In section \ref{methodology} the required methodology of the proposed approach is presented, starting with the robust UC with reserve constraint in \ref{adaptiveRUC}. Then the SFR model is presented in \ref{SFRmodel}, which takes the UC solutions as input. The outputs of the SFR model are used as the training data set for the LR model. How the LR works, and how the LR constraint is obtained is presented in \ref{LRmodel}. The adaptive robust UC formulation with LR can be found \ref{adaptiveLR}. The results are demonstrated in \ref{results}, and conclusions are drawn in \ref{conclution}.
\section{Methodology}
\label{methodology}
 This section presents the methodology. The main argument for using LR is that instead of trying to methodically linearize highly non-linear terms, it is possible to use historic data to represent frequency metrics with a line that is a function of system dynamic constants, available inertia, available reserve, and the amount of lost power. Such a procedure does not jeopardize accuracy through linearization and does not introduce unnecessary complexity and computational burden. The methodology is valid for active power unbalances in general, including generation outages. The proposed method tries to distinguish between outages that potentially violate tolerable frequency levels and the ones that do not. This type of problem can be dealt with dichotomous classification approaches like LR. The first step is to build a comprehensive set of data to train an accurate constraint. An adaptive robust UC with reserve constraint is used in this paper to obtain this data-set, which is explained in \ref{adaptiveRUC}. The UC problem is solved for different levels of reserve requirement, and ED is solved for all of the stochastic scenarios. The obtained results predominantly picture the possible feasible solutions that might be encountered in real-time. Using these data dynamic simulations are carried out to see the quality of frequency response in case of all potential outages. To perform the dynamic simulations an SFR model with no UFLS scheme is used (\ref{SFRmodel}). As the inputs of the SFR model have different levels of reserve and the amount of inertia is ignored, the simulation results will be a broad-ranging mix of tolerable frequency responses, poor responses, and even unstable cases. Analyzing the correlation between inputs and outputs of the SFR model facilitates the training of the LR model (\ref{LRmodel}), so it can distinguish the tolerable cases and the ones which will lead to poor frequency responses in case of outages. The obtained estimation of the LR model is further used in an adaptive robust UC formulation as an alternative constraint instead of the current reserve constraint (\ref{adaptiveLR}). Such formulation is inherently equipped with a constraint that is able to control the quality of frequency response of potential outages.
\subsection{Adaptive Robust UC with Reserve Constraint}\label{adaptiveRUC}
The Unit Commitment (UC) problem is a mixed-integer problem and is usually solved with MIL Programming solvers after the linearization of nonlinear terms. To solve the UC problem with uncertainty, an adaptive robust formulation is employed in \cite{bertsimas2012adaptive} and \cite{morales2018robust}. The formulation is robust, because it considers all of the possible realizations of the uncertain input, and makes sure that the chosen commitment status of the units, which is decided at the master level, will be feasible for any realization of the uncertain variable. The formulation is adaptive because the subproblem level is a function of the uncertain variables and can adapt the master level decision variable, depending on the different realizations of the uncertain variable. A general representation of UC problem with reserve constraint and uncertain wind power injection is provided here,
\begin{subequations} \label{e7}
\begin{equation}
{\min\limits_{x,p(w)} suc(x_{t,i})+gc(p_{i,t})}\tag{\ref{e7}}
\end{equation}
\begin{equation}
x_{t,i}-x_{t-1,i}=y_{t,i}-z_{t,i}\;\;\;\text{\scriptsize $t\in\mathcal{T},\;i\in\mathcal{I}$}\label{e8a}
\end{equation}
\begin{equation}
{y_{i,t}+z_{i,t}\leq1\;\;\;\text{\scriptsize $t\in\mathcal{T},\;i\in\mathcal{I}$}}\label{e8b}
\end{equation}
\begin{equation}
\sum\limits_{tt=t-UT_i+1}^{t}y_{tt,i}\leq x_{t,i}\;\;\;\text{\scriptsize $t\in\{UT_i,\dots,\mathcal{T}\}$}\label{e8c}
\end{equation}
\begin{equation}
\sum\limits_{tt=t-DT_i+1}^{t}z_{tt,i}\leq 1-x_{t,i}\;\;\;\text{\scriptsize $t\in\{UT_i,\dots,\mathcal{T}\}$}\label{e8d}
\end{equation}
\begin{equation}
{p(w)_{t,i}\geq \underline{\mathcal{P}_i}.x_{t,i}\;\;\;\text{\scriptsize $t\in\mathcal{T},\;i\in\mathcal{I},\;w\in\mathcal{W}\;,\alpha$}}\label{e9a}
\end{equation}
\begin{equation}
{p(w)_{t,i}+r(w)_{t,i}\leq \overline{\mathcal{P}_i}.x_{t,i}\;\;\;\text{\scriptsize $t\in\mathcal{T},\;i\in\mathcal{I},\;w\in\mathcal{W},\;\beta$}}\label{e9b}
\end{equation}
\begin{equation}
{p(w)_{t-1,i}-p(w)_{t,i}\leq \underline{\mathcal{R}_i}\;\;\;\text{\scriptsize $t\in\mathcal{T},\;i\in\mathcal{I},\;w\in\mathcal{W},\;\gamma$}}\label{e9c}
\end{equation}
\begin{equation}
{p(w)_{t,i}-p(w)_{t-1,i}\leq \overline{\mathcal{R}_i}\;\;\;\text{\scriptsize $t\in\mathcal{T},\;i\in\mathcal{I},\;w\in\mathcal{W},\;\delta$}}\label{e9d}
\end{equation}
\begin{equation}
{\sum\limits^{i\in\mathcal{I}}\big(p(w)_{t,i}\big)+wg(w)_t=d_t\;\;\;\text{\scriptsize $t\in\mathcal{T},\;w\in\mathcal{W},\;\zeta$}}\label{e9e}
\end{equation}
\begin{equation}
{wg(w)_{t}\leq w_t\;\;\;\text{\scriptsize $t\in\mathcal{T},\;w\in\mathcal{W}\;,\eta$}}\label{e9f}
\end{equation}
\begin{equation}
{\sum\limits_{ii\neq i}^{ii\in\mathcal{I}}\Big(\overline{\mathcal{P}}_i-p(w)_{t,ii}\Big)\geq p(w)_{t,i}\;\text{\scriptsize $t\in\mathcal{T},i\in\mathcal{I},w\in\mathcal{W},\mu$}}\label{e9g}
\end{equation}
\end{subequations}
The aim is to solve (\ref{e7}) subject to (\ref{e8a})-(\ref{e8d}), which only depends on binary variables, and (\ref{e9a})-(\ref{e9g}), which depend on both binary and real variables. $gc(.)$ is usually a quadratic cost function, which will be piece-wise linearized to be utilized in a MIL problem.  (\ref{e8a}) and (\ref{e8b}) represent the binary logic of the UC problem. (\ref{e8c}) and (\ref{e8d}) are the minimum up-time and minimum downtime constraints of the units. (\ref{e9a}) is the minimum power generation constraint, with dual multiplier $\alpha$. (\ref{e9b}) is the maximum power generation constraint with dual multiplier $\beta$, and states that the summation of power generation and power reserve of every online unit, should be less than the maximum output of the unit. (\ref{e9c}) and (\ref{e9d}) are ramp-down and ramp-up constraints, with dual multipliers $\gamma$ and $\delta$ respectively. (\ref{e9e}) is the power balance equation with dual multiplier $\zeta$. (\ref{e9f}) with dual multiplier, $\eta$ makes sure that the scheduled wind power is always equal to or less than the uncertain forecasted wind. (\ref{e9g}) is the current reserve constraint with dual multiplier $\mu$, and makes sure that in case of any contingency, there is enough headroom to compensate for lost generation.
Note that all the decision variables from (\ref{e9a}) to (\ref{e9g}) are a function of uncertain wind power realization. In practice, an iterative delayed constraint generating Benders' decomposition algorithm is used to solve this problem \cite{Taskin2011}. The problem is broken to a master problem minimization subjected to (\ref{e8a}) to (\ref{e8d}), and a subproblem with max-min form subjected to (\ref{e9a}) to (\ref{e9g}).
\begin{equation} \label{e10}
{\min\limits_{x}\Big( suc(x_{t,i})+\max\limits_{w\in\mathcal{W}}\min\limits_{p} gc(p_{i,t})\Big)}
\end{equation}
The minimization on the master level is subjected to (\ref{e8a}) to (\ref{e8d}), and the subproblem level minimization is subjected to (\ref{e9a}) to (\ref{e9g}). The subproblem minimization problem determines the ED cost for a fixed commitment $\hat{x}$, and then it's maximized over the uncertainty set $\mathcal{W}$. Here the concept of duality in linear problems can be used. As the strong duality suggests, \emph{the dual has an optimal solution if and only if the primal does, and the solutions are equal}. Taking the dual of subproblem converts the max-min form into a maximization problem. Considering the decomposed form of the problem, the feasible region of subproblem maximization is independent of $x$. So the subproblem maximization can be described as a set of extreme points and extreme rays of solution region. Let $\mathcal{O}$ be the complete set of possible extreme points, and $\mathcal{F}$ be the complete set of possible extreme rays. These properties will later be used to define the decomposed master problem. In the iterative solution process, the binary variable, $\hat{x}_{t,i}$, is obtained form masters' problem, hence it's fixed. With that in mind, and defining the auxiliary variable $\phi$, as an understimator of optimal subproblem objective value, the dual form of subproblem is defined as follows,
\begin{subequations}\label{e11obj}
\begin{equation}
\phi\geq \max\limits_{p}{\left(\begin{aligned}
   &\sum\limits^{t\in\mathcal{T}}\sum\limits^{i\in\mathcal{I}}\alpha_{t,i}(\underline{\mathcal{P}}_i.\hat{x}_{t,i})\\
    -&\sum\limits^{t\in\mathcal{T}}\sum\limits^{i\in\mathcal{I}}\beta_{t,i}(\overline{\mathcal{P}}_i.\hat{x}_{t,i})\\
    -&\sum\limits^{t\in\mathcal{T}}\sum\limits^{i\in\mathcal{I}}(\gamma_{t,i}.\underline{\mathcal{R}}_i+\delta_{t,i}.\overline{\mathcal{R}}_i)\\
    -&\sum\limits^{t\in\mathcal{T}}(\zeta_{t}.d_t+\eta_{t}.w_t)\\
    -&\sum\limits^{t\in\mathcal{T}}\sum\limits^{i\in\mathcal{I}}\mu_{t,i}\Big(\sum\limits_{ii\neq i}^{ii\in\mathcal{I}}\big(\overline{\mathcal{P}}_i\big)\Big)
\end{aligned}\right)} \tag{\ref{e11obj}}
\end{equation}
\begin{equation}
\begin{aligned}
&C_i-\alpha_{t,i}+\beta_{t,i}+\gamma_{t,i}+\delta_{t,i}+\zeta_{t}+\\
&\eta_{t}+\sum\limits_{ii}^{ii\in\mathcal{I}}\mu_{t,ii} \geq 0 \;\;\;\text{\scriptsize $t\in\mathcal{T},\;i\in\mathcal{I}$} \label{e11a}
\end{aligned}
\end{equation}
\begin{equation}
\zeta_{t} + \eta_{t} \geq 0 \;\;\;\text{\scriptsize $t\in\mathcal{T}$}  \label{e11b}
\end{equation}
\begin{equation}
\alpha,\beta,\gamma,\delta,\eta,\mu \geq 0 \text{ and $\zeta$ is free}  \label{e11c}
\end{equation}
\end{subequations}
The dual form is (\ref{e11obj}) subject to (\ref{e11a}) to (\ref{e11c}). $\zeta$ is a free variable, because (\ref{e9e}) is an equality. To find out more about writing a standard form of a problem, and taking the dual, have a look at \cite{lahaie2008take}. The term $\eta_t w_t$ in the dual objective function is nonlinear, so an outer approximation approach \cite{fletcher1994solving} is employed to cope with it. The objective function of subproblem dual is a function of all dual variables and fixed $\hat{x}_{t,i}$ from the master problem in previous iteration. Let's define the set of dual variables as $u$ and the dual objective solution as $f(\hat{x}_{t,i},\hat{u})$. Then the master problem is defined as follows,
\begin{equation} \label{emaster}
\begin{split}
\min\limits_{x}\quad & suc(x_{t,i})+\phi,\\
\mathbf{s.t.}\quad  & (\ref{e8a})\; \text{to}\; (\ref{e8d}),\\
& \phi\geq{f(\hat{x}_{t,i},\hat{u})}\;\;\forall{u}\in\mathcal{O}\\
& 0\geq{f(\hat{x}_{t,i},\hat{u})}\;\;\forall{u}\in\mathcal{F}\\
\end{split}
\end{equation}
Iterative solution process starts with empty sets of $\mathcal{O}$ and $\mathcal{F}$. Then if the subproblem solution corresponding to $\hat{x}_{t,i}$ ($f(\hat{x}_{t,i},\hat{u})$), is feasible, an optimality cut is generated and added to $\mathcal{O'}$. And if the subproblem solution corresponding to the $\hat{x}_{t,i}$ is infeasible, $f(\hat{x}_{t,i},\hat{u})$ is unbounded and a feasibility cut is generated and added to $\mathcal{F'}$. The iterations continue until $\phi$ is converged enough. The iterative algorithm is presented in algorithm \ref{algo1}.
\begin{algorithm}[t!]
\caption{Iterative UC with reserve}
\label{algo1}
\textbf{Input}: System specifications, wind uncertainty set, power demand\\
\textbf{Output}: $\epsilon$-optimal solution
\begin{algorithmic}[1]
    \State $j = 0$
    \While {$\lvert\phi^j(\hat{x}^j_{t,i},\hat{u}^{j})-\phi^j(\hat{x}^j_{t,i},\hat{u}^{j-1})\rvert<\epsilon$}
        \State \parbox[t]{\dimexpr\linewidth-\algorithmicindent}{%
        Solve master, minimizing $suc(x^j_{t,i})+\phi^j(x_{t,i}^j,\hat{u}^{j-1})$ to get $\hat{x}_{t,i}^j$}
        \State \parbox[t]{\dimexpr\linewidth-\algorithmicindent}{%
        Solve subproblem using outer approximation, maximizing $f(\hat{x}^j_{t,i},u^j)$ to get $\hat{u}^j$}
        \State If $f(\hat{x}^j_{t,i},\hat{u}^j)$ is bounded $\rightarrow \mathcal{O'}\cup \{\hat{u}_j \}$
        \State If $f(\hat{x}^j_{t,i},\hat{u}^j)$ is unbounded $\rightarrow \mathcal{F'}\cup \{\hat{u}_j \}$
        \State $j=j+1$
    \EndWhile
\end{algorithmic}
\end{algorithm}
The UC problem is solved for different levels of reserve requirement. The optimal commitment variables are then used to solve the ED problem for various stochastic wind scenarios to build an initial data-set, which will be implemented in the SFR model.
\subsection{System Frequency Response (SFR) Model}\label{SFRmodel}
\begin{figure}[t]
\centering
\includegraphics[width=0.9\linewidth]{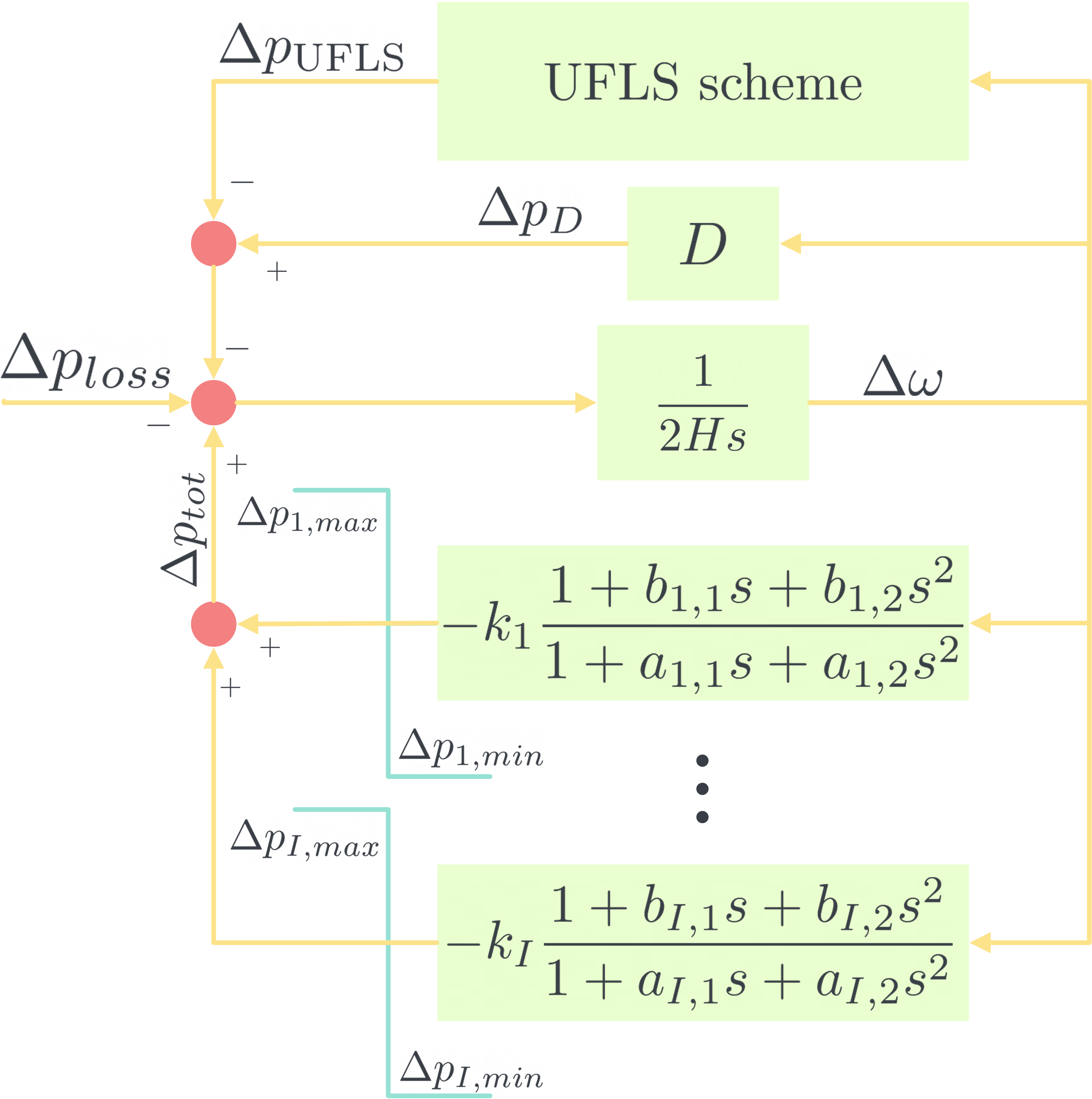}
\caption{SFR model.}
\label{fSFRmodel}\end{figure}

This section briefly presents SFR models used to analyze the frequency stability of small isolated power systems. The model is able to reflect the underlying short-term frequency response of small isolated power systems. Figure \ref{fSFRmodel} details the power-system model typically used to design UFLS schemes for an island power system, consisting of $\mathcal{I}$ generating units.
Each generating unit $i$ is represented by a second-order model approximation of its turbine-governor system. In fact, dynamic frequency responses are dominated by rotor and turbine-governor system dynamics. Excitation and generator transients can be neglected for being much faster than the turbine-governor dynamics. Since frequency can be considered uniform, an equivalent normalized system inertia $\widetilde{\mathcal{H}}$ can be defined as follows,
\begin{equation}
   \widetilde{\mathcal{H}}= \sum\limits^{i\in\mathcal{I}}  \frac{\mathcal{H}_i\mathcal{M}^{base}_i}{\mathcal{S}^{base}}
\label{e17}\end{equation}
The overall response of loads can be considered by means of a load-damping factor $D$ if its value is known. The gain $k_i$ and parameters $a_{i,1}$, $a_{i,2}$, $b_{i,1}$ and $b_{i,2}$, of each generating unit $i$ can be deduced from more accurate models or field tests. Since primary spinning reserve is finite, power output limitations $\Delta p_{i,min}$ and $\Delta p_{i,max}$ are forced. So the units can only participate as much as their available reserve. The complete model is explained in \cite{sigrist2016island}. 
In practice, the UFLS scheme is designed to stabilize the system after large outages. For the purpose of building a data set to train the LR model, the UFLS scheme should be deactivated so the results capture the free frequency responses, including the ones that lead to instability quantified by unacceptable low-frequency nadir and steady-state frequency. Note however that the UFLS scheme will be considered to quantify the expected amount of UFLS when comparing the new reserve constraints with the current one in section \ref{results}.
\subsection{Logistic Regression (LR)}\label{LRmodel}
\begin{figure}[t]
\centering
\includegraphics[width=1\linewidth]{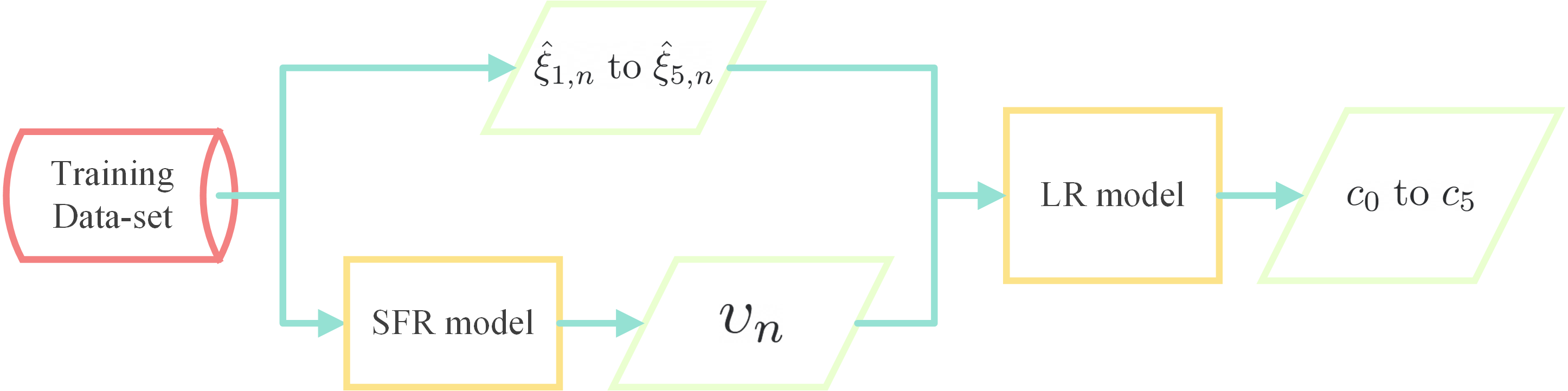}
\caption{Flowchart of calculating LR coefficients.}
\label{eq9Fig}\end{figure}

Regression methods are used for data analysis, concerned with describing the relationship between a response variable and one or more explanatory variables. Sometimes the output variable needs to be discrete, taking one or more possible values. In these instances, logistic regression is usually used. Consider a collection of $m$ independent variables denoted by the vector $\xi^{'} =(\xi_1,\xi_2,\dots,\xi_m)$ related to a dichotomous dependent variable $\upsilon$, where $\upsilon$ is typically coded as 1 or 0 for its two possible categories. Considering that for a $(0,1)$ random variable, the expected value of $\upsilon$ is equal to the probability of $\upsilon=1$ (i.e., $\pi (\upsilon=1)$), and is defined here,
\begin{equation}
    \pi(\upsilon=1)=\frac{1}{1+e^{-(c_0+c_1\xi_1+c_2\xi_2+\dots+c_m\xi_m)}}
\label{e20}\end{equation}
The regression coefficients $c_0$ to $c_m$ in the logistic model (\ref{e20}) provide important information about the relationships of the independent variables in the model to the dichotomous dependent variable. For the logistic model, these coefficients are used to estimate the odds ratio. Odds are defined as the ratio of the probability that some event will occur divided by the probability that the same event will not occur. Thus the odds for the event $\upsilon=1$ is,
\begin{equation}
    odds(\upsilon=1)=\frac{\pi(\upsilon=1)}{1-\pi(\upsilon=1)}
\label{e21}\end{equation}
Generally the conditional probability that the outcome presents is denoted by $\pi (\upsilon)$. The logit transformation of the probability $\pi(\upsilon=1)$ is defined as natural logarithm of the odds of event $\upsilon=1$, and considering (\ref{e20}) is defined as,
\begin{equation}
    \begin{split}
        logit\big(\pi(\upsilon=1)\big)&=ln\Big(\frac{\pi(\upsilon=1)}{1-\pi(\upsilon=1)}\Big)\\&=
    c_0+c_1\xi_1+c_2\xi_2+\dots+c_m\xi_m
    \end{split}
\label{e22}\end{equation}
This is the \emph{logit form} of the model and is given by a linear function \cite{kleinbaum2013applied}. The logit transformation is primarily applied to convert a variable that is bounded by 0 and 1 (i.e., probabilities) to a variable with no bounds \cite{huang2013odds}. When $logit\big(\pi(\upsilon=1)\big)$ goes toward $+\infty$, the probability of event $\upsilon=1$ gets closer to $1$, and when $logit\big(\pi(\upsilon=1)\big)$ goes toward $-\infty$, the probability of event $\upsilon=1$ gets closer to $0$. Usually $logit\big(\pi(\upsilon=1)\big)=0$ is considered as a cut-point, that separates those events with the probability of more than $0.5$ on the positive side, and those events with the probability of less than $0.5$ on the negative side. Depending on the required accuracy for the model, different cut-points can be chosen.\\
As the frequency response of the system after contingencies is highly nonlinear, different approaches are employed in the literature to somehow linearize and include them in the UC problem. Some of these approaches are mathematically complicated and often tremendously burdensome for the solver. The argument here is that instead of linearizing the complex nonlinear equations, the output of developed SFR models can be analyzed to drive a linear constraint. To do so, the frequency response after each contingency can be marked as acceptable or unacceptable, depending on whether it violates the predefined limits or not. Then logistic regression is employed here to analyze the data and separate acceptable and unacceptable results with a trained line. This line is added later to the UC problem as a constraint by replacing the current reserve constraint (\ref{e9g}). Such constraint can improve the frequency response quality and reduce the amount of load shedding due to unexpected outages, as it takes into account the expected dynamic behavior of the system.\\
As it's going to be further discussed in the results section, the independent variables that are considered in the LR model are the weighted summation of online inertia ($\xi_1$), the summation of inverse droop of the online units ($\xi_2$), lost power ($\xi_3$), lost power divided by the corresponding demand of that hour ($\xi_4$), and remaining of the reserve power after generator outages ($\xi_5$). An input data set of different UC solutions will be used to calculate the independent variables ($\hat{\xi}_{1,n}$ to $\hat{\xi}_{5,n}$) for every possible generator outage $n$, and then the result is fed into the LR model, to obtain coefficients $c_0$ to $c_5$. How each incident is assigned with a dependent variable ($\upsilon_n$), by carrying out the dynamic simulations, is later explained. The general procedure is shown in figure \ref{eq9Fig}. 
The general form of the trained constraint estimated by LR procedure is presented as follows,\\
\begin{equation}
\begin{split}
    &c_0+c_1\Big(\sum\limits_{ii\neq i}^{ii\in\mathcal{I}}\mathcal{H}_{ii}\mathcal{M}_{ii}^{base}x_{t,ii}\Big)+\\&c_2\Big(\sum\limits_{ii\neq i}^{ii\in\mathcal{I}}\mathcal{K}_{ii}x_{t,ii}\Big)+c_3 p_{t,i}+\frac{c_4}{d_t}p_{t,i}+\\&c_5\Big(\sum\limits_{ii\neq i}^{ii\in\mathcal{I}}(\overline{\mathcal{P}_{ii}}x_{t,ii}-p_{t,ii})\Big)\geq\psi\;\;\;\;\text{\scriptsize $t\in\mathcal{T},\;i\in\mathcal{I}$}
\end{split}
\label{e23}\end{equation}
This constraint enables the UC problem to also take into account the inertia and time constants of the system. The purpose is to improve the quality of frequency response with these measures.
\subsection{Adaptive Robust UC with LR constraint}\label{adaptiveLR}
\begin{figure}[t]
\centering
\includegraphics[width=\linewidth]{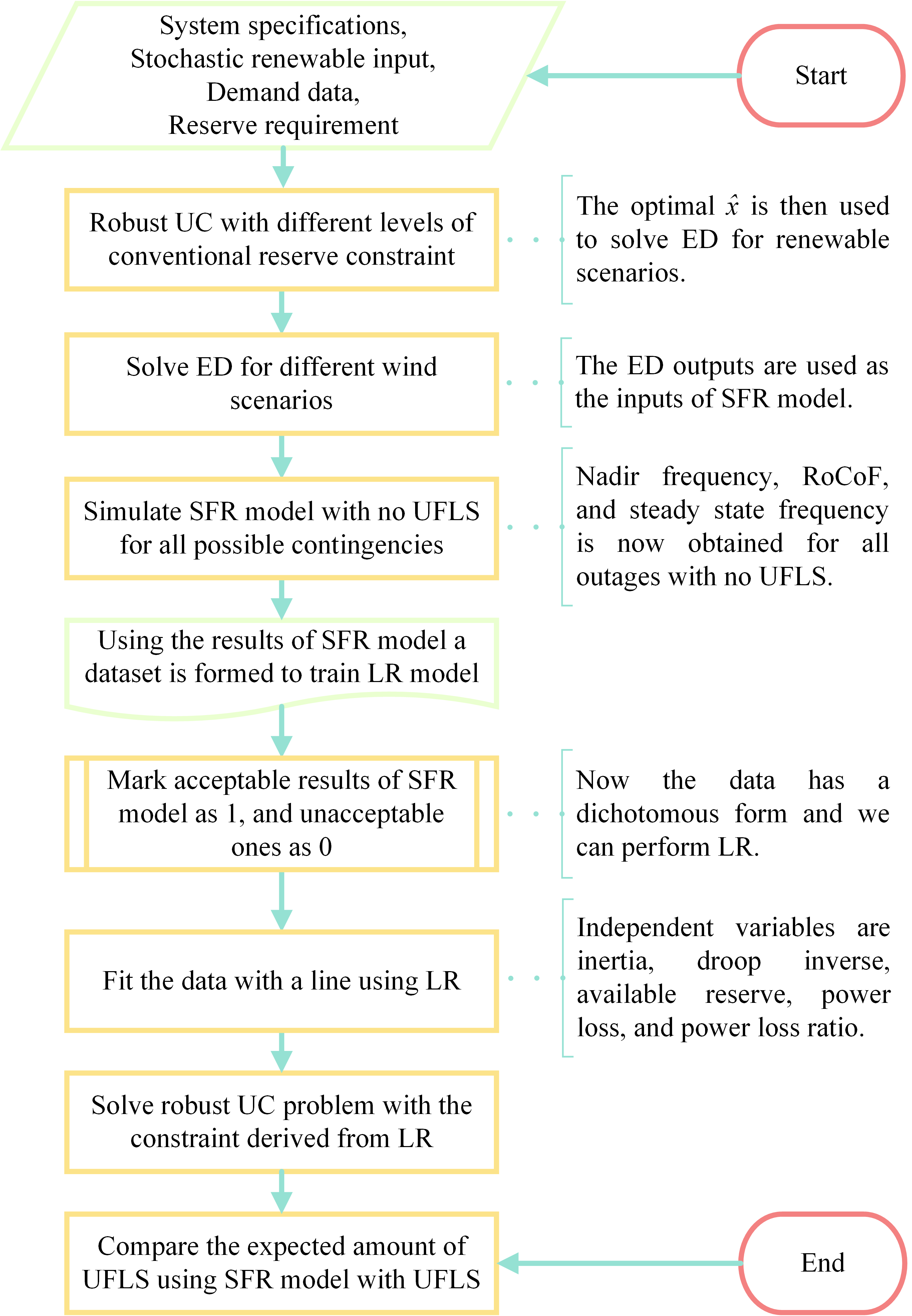}
\caption{Flowchart of the proposed method.}
\label{fflowchart}\end{figure}

The general formulation is similar to \ref{e7}, but reserve constraint in \ref{e9g} is replaced by the LR constraint in \ref{e23}. The subproblem dual with the new constraint will become as follows,
\begin{subequations}\label{RUCwithLRobj}
\begin{equation}
\phi\geq \max\limits_{p}{\left(\begin{aligned}
   &\sum\limits^{t\in\mathcal{T}}\sum\limits^{i\in\mathcal{I}}\alpha_{t,i}(\underline{\mathcal{P}}_i.\hat{x}_{t,i})\\
    -&\sum\limits^{t\in\mathcal{T}}\sum\limits^{i\in\mathcal{I}}\beta_{t,i}(\overline{\mathcal{P}}_i.\hat{x}_{t,i})\\
    -&\sum\limits^{t\in\mathcal{T}}\sum\limits^{i\in\mathcal{I}}(\gamma_{t,i}.\underline{\mathcal{R}}_i+\delta_{t,i}.\overline{\mathcal{R}}_i)\\
    -&\sum\limits^{t\in\mathcal{T}}(\zeta_{t}.d_t+\eta_{t}.w_t)\\
    -&\sum\limits^{t\in\mathcal{T}}\sum\limits^{i\in\mathcal{I}}\rho_{t,i}\Big(c_0\\&+c_1\big(\sum\limits_{ii\neq i}^{ii\in\mathcal{I}}\mathcal{H}_{ii}\mathcal{M}_{ii}^{base}x_{t,ii}\big)\\&+c_2\big(\sum\limits_{ii\neq i}^{ii\in\mathcal{I}}\mathcal{K}_{ii}x_{t,ii}\big)\\&+c_5\big(\sum\limits_{ii\neq i}^{ii\in\mathcal{I}}(\overline{\mathcal{P}_{ii}}x_{t,ii})\big)\Big)
\end{aligned}\right)} \tag{\ref{RUCwithLRobj}}
\end{equation}
\begin{equation}
\begin{aligned}
&C_i-\alpha_{t,i}+\beta_{t,i}+\gamma_{t,i}+\delta_{t,i}+\zeta_{t}+\eta_{t}+\\&
(c_3+\frac{c_4}{d_t})\rho_{t,i}+c_5\sum\limits_{ii\neq{i}}^{ii\in\mathcal{I}}\rho_{t,ii} \geq 0 \;\;\;\text{\scriptsize $t\in\mathcal{T},\;i\in\mathcal{I}$} \label{RUCwithLRobja}
\end{aligned}
\end{equation}
\begin{equation}
\zeta_{t} + \eta_{t} \geq 0 \;\;\;\text{\scriptsize $t\in\mathcal{T}$}  \label{RUCwithLRobjb}
\end{equation}
\begin{equation}
\alpha,\beta,\gamma,\delta,\eta,\rho \geq 0 \text{ and $\zeta$ is free}  \label{RUCwithLRobjc}
\end{equation}
\end{subequations}
As the objective function in the primal form and all the constraints that only involve binary variables are the same, the master problem remains the same as section \ref{adaptiveRUC}. The iterative solution procedure here is the same as algorithm \ref{algo1}. A flowchart of the different steps of the proposed method is presented in figure \ref{fflowchart}.
\section{Results}\label{results}
\subsection{Case study and inputs}
\begin{figure}[t]
    \centering
    \includegraphics[width=\linewidth]{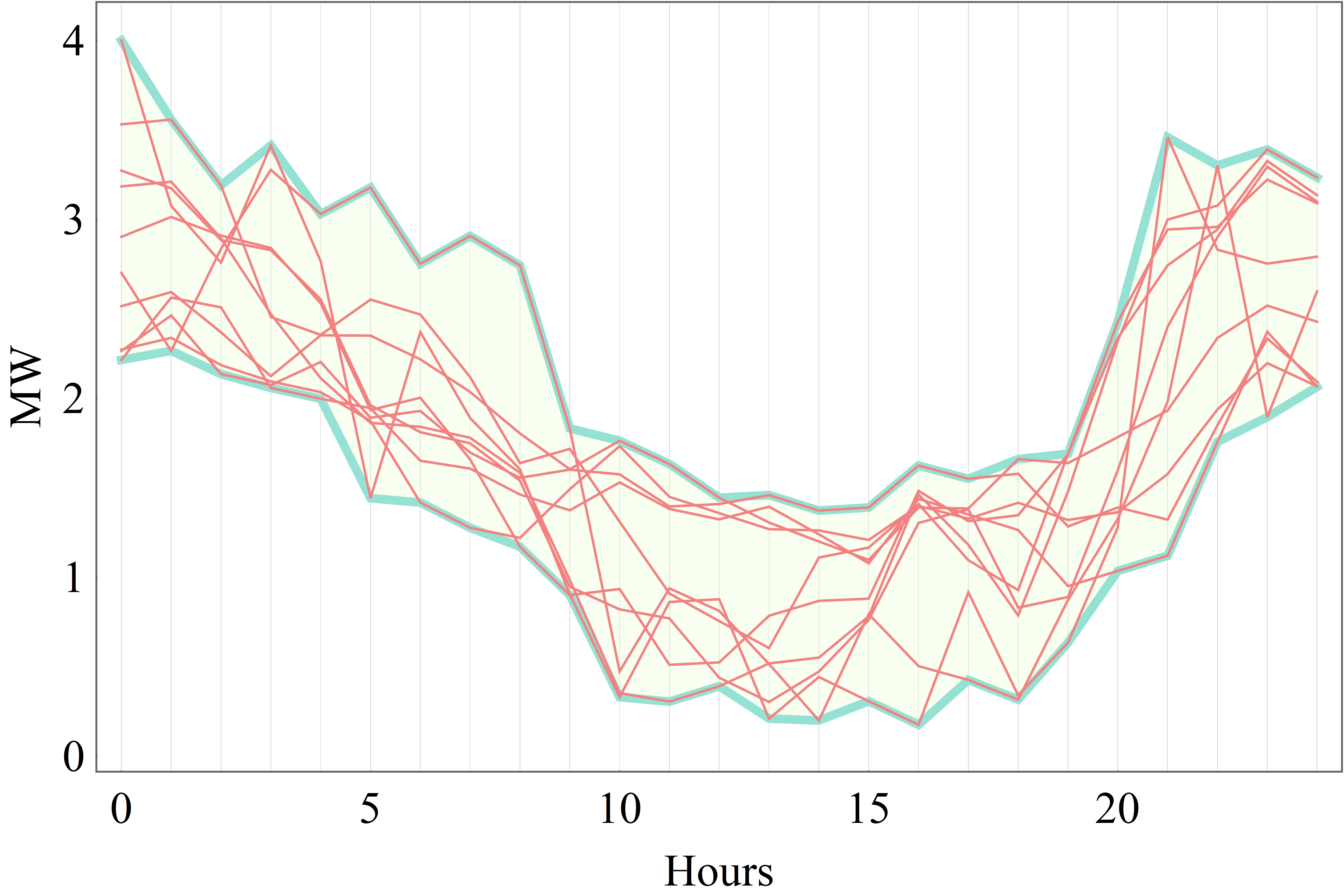}
    \caption{Wind data uncertainty set}
    \label{fwindData}
\end{figure}

Simulations for the proposed methodology are carried on the real power system of the island of La Palma island, one of Spain's Canary Islands. The yearly demand in 2018 is reported about 277.8 GWh (average hourly demand of 31.7 MWh), supplied by eleven Diesel generators pre-dominantly. According to \cite{de2019consejeria}, the installed capacity of the La Palma island power system mounts to 117.7 MW, where about $6\%$ of the installed capacity belongs to wind power generation. RES covers about $10\%$ of the yearly demand. The input data for solving the UC problem is obtained from real data. Different scenarios of forecasted wind generation data of a sample day are chosen, which also provide the upper bound and the lower bound of the robust formulation. Wind data with 10 scenarios is shown in figure \ref{fwindData}.\\
An initial data set is required to train the LR model. A complete data-set that includes different reserve levels and different wind levels is preferred, providing enough information for the LR model, so it can reliably distinguish acceptable and unacceptable results. To achieve such a training data set, the conventional day-ahead robust UC is solved for ascending reserve requirements levels, starting from zero requirement until the problem becomes infeasible. In the conventional UC, the reserve requirement is typically the largest generation source under moderate RES penetration. A multiplier is defined here for the reserve requirement starting from 0, with 0.1 ascending steps, until 1.5, which is the point that problem becomes infeasible in this case. Then the ED solution of 10 wind scenarios for each reserve requirement level is fed to the SFR model, and all single generator outages are simulated. Obtained results confirm that other system characteristics, like online inertia, lost power, lost power percentage, and the droop of the turbine-governor system is very related to the quality of the frequency response, and to a larger extent than the amount of reserve. Table \ref{tcorrelation} shows the Pearson's correlation between mentioned characteristics and frequency response metrics, for more than 20000 single outages, simulated by the SFR model.
\begin{table}[tp]
\caption{Pearson's correlation between parameters}
\label{tcorrelation}
\centering
\begin{tabular}{p{0.2\linewidth}|p{0.2\linewidth}p{0.2\linewidth}p{0.2\linewidth}}
\toprule
& \centering $f^{nadir}$ & \centering $f^{qss}$ & \centering $RoCoF$ \tabularnewline
\hline
\rowcolor{myTeal}
\centering $\sum\mathcal{H}$& \centering 0.568 & \centering 0.558 & \centering 0.668 \tabularnewline
\centering $\sum\mathcal{K}$& \centering 0.286 & \centering 0.283 & \centering 0.319 \tabularnewline
\rowcolor{myTeal}
\centering $p^{loss}$& \centering -0.561 & \centering -0.532 & \centering -0.876 \tabularnewline
\centering $\nicefrac{p^{loss}}{d}$& \centering -0.617 & \centering -0.588 & \centering -0.965 \tabularnewline
\rowcolor{myTeal}
\centering $\sum r$& \centering 0.506 & \centering 0.516 &  \centering 0.269 \tabularnewline
\bottomrule
\end{tabular}
\end{table}
As expected, the ratio of lost generation to hourly demand has the best correlation with frequency metrics, as the big outages relatively to the whole generation tend to disturb frequency considerably. Interestingly enough, the sum of available reserve has a weaker correlation with frequency metrics, compared to the others. Meaning that fulfilling reserve criteria does not guarantee the quality of frequency response in small power systems with low inertia, as the remaining units are not fast enough to compensate the power mismatch, while the frequency is dropping fast due to lack of inertia. So other parameters like total available inertia and power loss ratio are better representatives of the system dynamics.\\
La Palma island, like other power systems, is equipped with a UFLS scheme that sheds load depending on the severity of RoCoF (Rate of Change of Frequency) and frequency deviation. Large generation outages lead to UFLS since primary frequency response is insufficient. The current practice of power schedule in islands only includes the reserve criteria to cover outages, and as mentioned, reserve does not have a strong correlation with the frequency response metrics. So improvement in frequency response quality is expected by including more correlated parameters in UC formulation, leading to less UFLS.\\
Using the obtained data-set from the SFR model, a dichotomous data-set is formed. The simulation results for all single outages are classified into two groups, which eventually will be treated as the dependant variable in the LR model. For the purpose of this paper, any generator outage incident which leads to frequency nadir less than 47.5 Hz, or a RoCoF less than $-0.5$ Hz/s, or steady-state frequency less than 49.6 Hz, is considered as an unacceptable incident and is assigned 0. Other incidents are considered acceptable and are assigned as 1. To have an accurate LR model, all correlated parameters are included in the set of independent variables. Obtained LR coefficients for La Palma island are presented in table \ref{tcoefficients}.
\begin{table}[b]
\caption{Logistic regression coefficients}
\label{tcoefficients}
\centering
\begin{tabular}{>{\centering\arraybackslash}p{0.08\linewidth}>{\centering\arraybackslash}p{0.32\linewidth}|>{\centering\arraybackslash}p{0.1\linewidth}>{\centering\arraybackslash}p{0.3\linewidth}}
\toprule
\multicolumn{2}{c|}{Independent variable} & \multicolumn{2}{c}{LR coefficient}  \\
\hline
\rowcolor{myTeal}
\textendash &  \textendash  &  $c_0$ & 26.577 \\
$\xi_1$ & $\sum\limits_{ii\neq i}^{ii\in\mathcal{I}}\mathcal{H}_{ii}\mathcal{M}_{ii}^{base}x_{t,ii}$ & $c_1$ & -0.366 \\
\rowcolor{myTeal}
$\xi_2$& $\sum\limits_{ii\neq i}^{ii\in\mathcal{I}}\mathcal{K}_{ii}x_{t,ii}$ & $c_2$ & 0.102 \\
$\xi_3$& $p_{t,i}$ & $c_3$ & 1.484 \\
\rowcolor{myTeal}
$\xi_4$& $\nicefrac{p_{t,i}}{d_t}$ & $c_4$ & -173.995 \\
$\xi_5$& $\sum\limits_{ii\neq i}^{ii\in\mathcal{I}}(\overline{\mathcal{P}_{ii}}x_{t,ii}-p_{t,ii})$ & $c_5$ &  2.356 \\
\bottomrule
\end{tabular}
\end{table}
These coefficients can be implemented to (\ref{e23}), with an adjustable cut-point $\psi$ to set up a new constraint. As discussed in section \ref{LRmodel}, the logit form is a transformation of probabilities. In this case, incidents that are more probable to be acceptable should have a positive logit and a probability close to 1. On the other hand, incidents that are more probable to be unacceptable should have a negative logit and a probability close to 0. There will also be some errors, mainly around 0.5 probability, meaning that some acceptable incidents might end up possessing a negative logit value and vice versa. Depending on the preferred outcome, a proper cut-point can be chosen to create a more conservative or less conservative constraint.
\begin{figure}[htp]
    \centering
    \includegraphics[width=\linewidth]{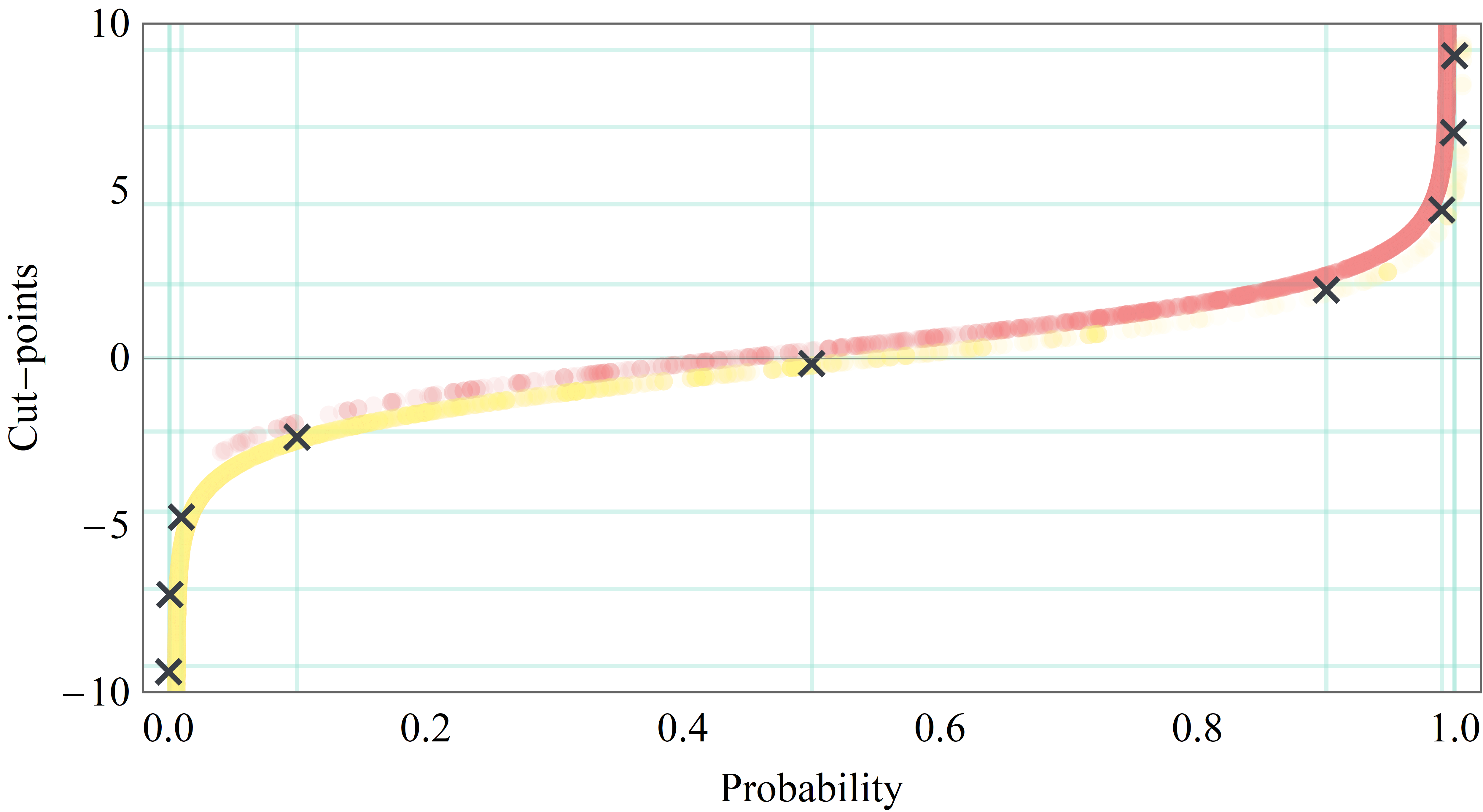}
    \caption{Logistic regression approximation}
    \label{lrapprox}
\end{figure}
\begin{figure}[tp]
    \centering
    \includegraphics[width=\linewidth]{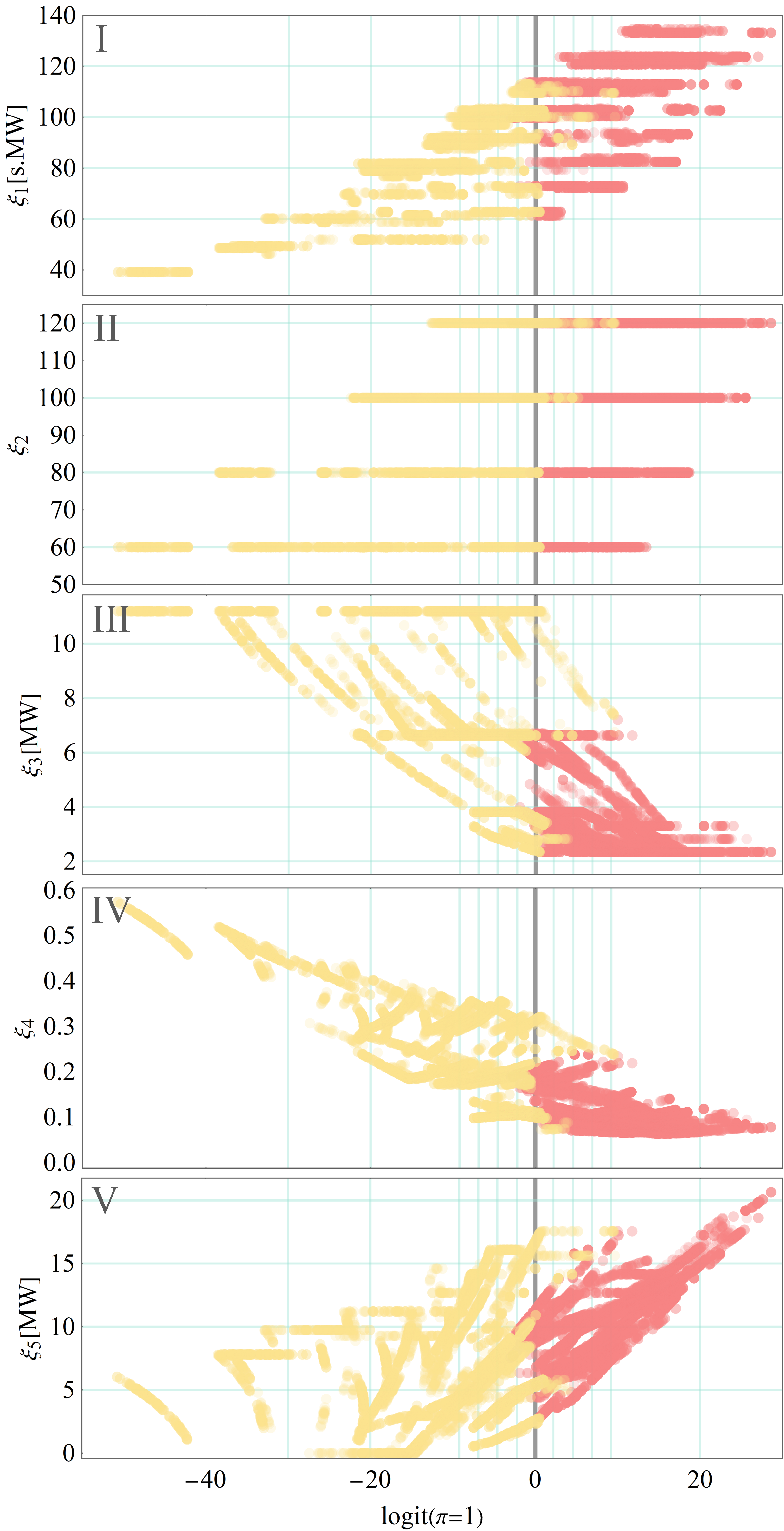}
    \caption{Logit transformation - variables}
    \label{flogitxi}
\end{figure}
Figure \ref{lrapprox} shows how accurate the applied logistic regression can separate acceptable and unacceptable results. Acceptable results are in red and unacceptable ones are in yellow.\\
Depending on the required conservativeness a cut-point is chosen. For example $\psi=0$ corresponds to $\pi(\upsilon=1)=0.5$. Putting $\psi=0$, means all the incidents that their probability of being unacceptable is more than 0.5, will be eliminated, hence it's very conservative. A less conservative approach is to only eliminate the instances with the probability of being unacceptable more than 0.9 ($\pi(\upsilon=1)\leq 0.1$). Then $\psi$ should be set equal to $-2.12$ (considering (\ref{e22})). Some probabilities and their corresponding cut-points are shown with the cross sign in figure \ref{lrapprox}.\\
In figure \ref{flogitxi}, it's shown how different independent variables, $\xi_1$ to $\xi_5$ (as defined in table \ref{tcoefficients}), are described by the logistic regression approximation. Those incidents that are marked as acceptable before are the red dots, and unacceptable incidents are the yellow dots.
There are some errors, especially close to $logit(\pi=1)$ line, but the overall accuracy is justifiable. The summation of online inertia, $\xi_1$, is depicted in the figure \ref{flogitxi}.I. Acceptable results are more concentrated on the top side which are the incidents with higher online inertia, and as the online inertia drops, the dots move towards unacceptable results. A similar conclusion can be drawn for the summation of the droops of online turbine-governor systems, $\xi_2$, shown in figure \ref{flogitxi}.II. The amount of lost generation, $\xi_3$, is depicted in figure \ref{flogitxi}.III. As expected, larger outages tend to result in unacceptable incidents and as the figure goes toward smaller outages, the concentration of acceptable incidents grows. The same conclusion is derived from figure \ref{flogitxi}.IV, which shows the ratio of lost generation to hourly demand, $\xi_4$. Available reserve is depicted in figure \ref{flogitxi}.V. Generally incidents with a higher amount of online reserve tend to lead to better results, but still there are a considerable number of incidents that lead to unacceptable results, although they have a relatively high available reserve. This confirms that the available reserve is not the best indicator to ensure the quality of dynamic response after outages. The goal is to improve the quality of frequency response by including all of these independent variables, each of them weighted carefully with logistic regression coefficients.
\subsection{Comparison of different methods}
\begin{figure}[t]
    \centering
    \includegraphics[width=\linewidth]{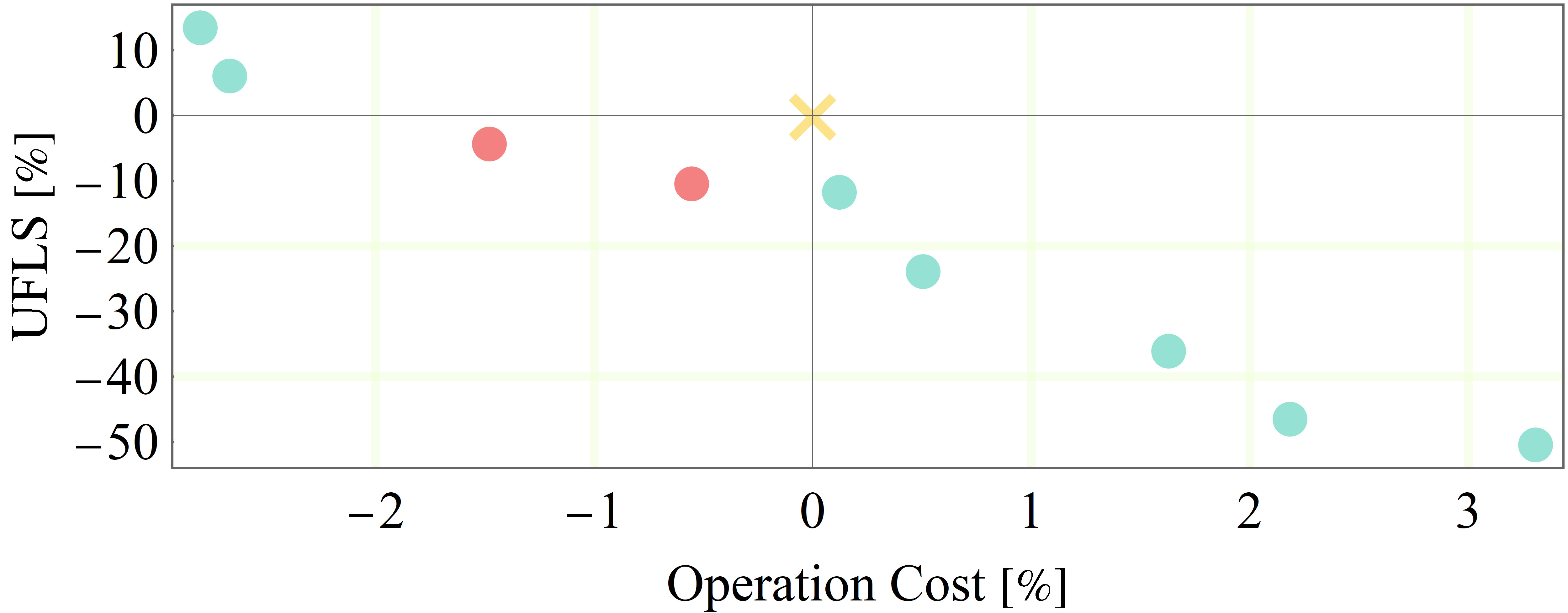}
    \caption{Average UFLS and operation cost in percentage}
    \label{ocufls}
\end{figure}

\begin{figure}[tp]
    \centering
    \includegraphics[width=\linewidth]{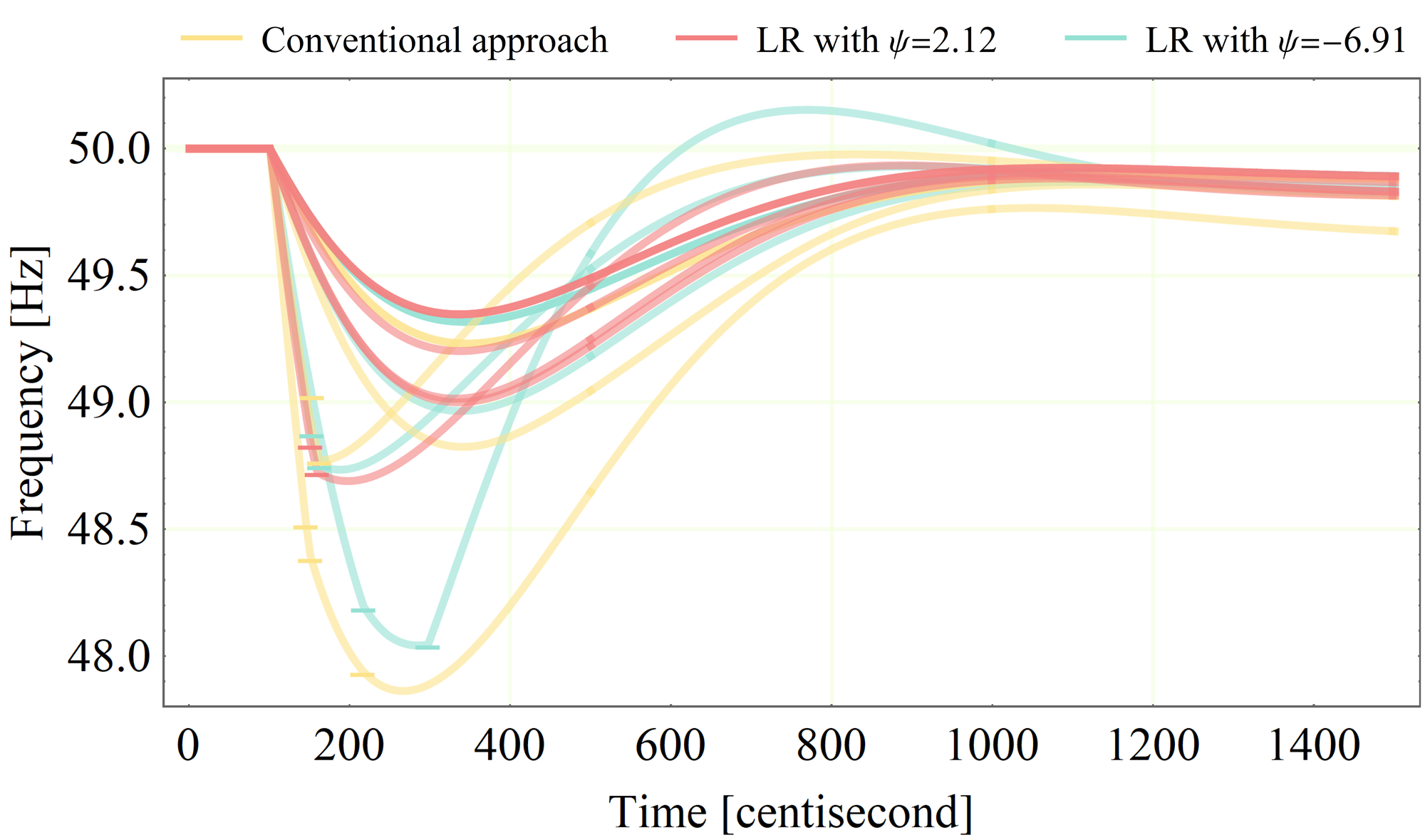}
    \caption{Frequency response after outages for a random hour with UFLS.} 
    \label{fwithufls}
\end{figure}

\begin{figure}[tp]
    \centering
    \includegraphics[width=\linewidth]{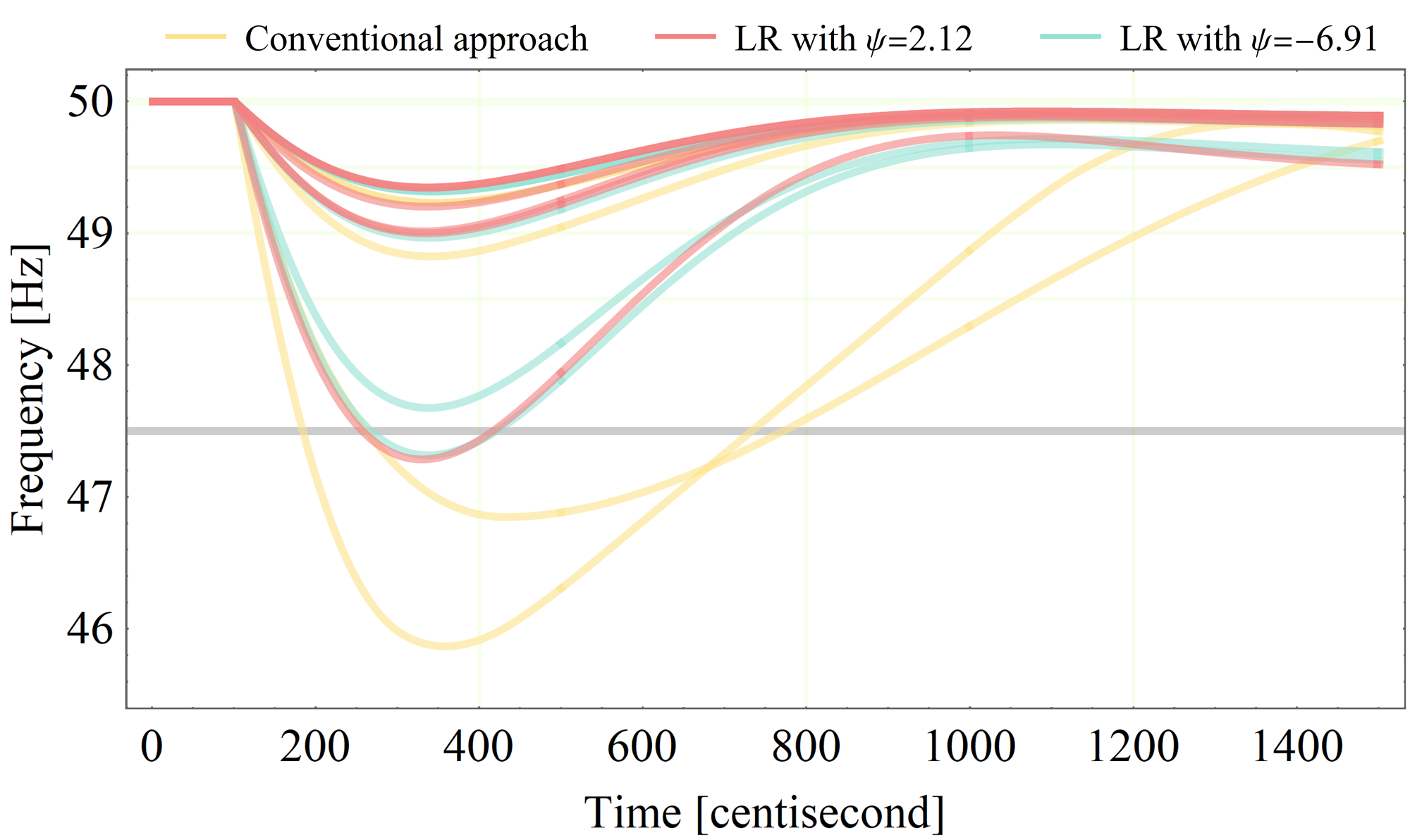}
    \caption{Frequency response after outages for a random hour without UFLS.}
    \label{fwithoutufls}
\end{figure}

Simulations are carried out for three different methods:\\
\textbf{Conventional approach:} the conventional formulation of robust UC, that the frequency response after outage is only guaranteed by reserve criteria. Reserve requirement is the biggest online generation infeed.\\
\textbf{LR:} the proposed logistic regression method. Reserve criteria is substituted with a constraint that is trained by LR model. Different cut-points ($\psi$) are considered to asses the effectiveness of the proposed method, when the LR constraint is looser (smaller $\psi$) or tighter (bigger $\psi$).\\
\textbf{OCT:} To also compare the proposed method with other recent data-driven methods in the literature, optimal classification trees are implemented to train a constraint, as introduced in \cite{lagos2021data}. The outputs of SFR model are classified in acceptable and unacceptable incidents, using the MIL solution method of \cite{Bertsimas2017}. As solving the optimization problem for classification becomes very hard with a big set of inputs and high depth of trees, only the biggest hourly outage of a limited number of scenarios is fed to OCT problem as input, with the maximal depth of one and two.\\
A comparison of frequency response indicators for conventional approach, LR, and OCT is presented in table \ref{tresults}. Frequency-response quality indicators are the average amount of UFLS which is obtained from SFR with UFLS active, average frequency nadir, average RoCoF, and average quasi-steady-state frequency, which is obtained from SFR with UFLS deactivated. The changes in average UFLS and operation costs relative to the conventional approach are presented in percentage too. Cut-points beyond $\psi= 2.12$ make the problem infeasible, so $\psi=2.12$ is presented in the table as the most conservative cut-point that is feasible. The results assert that more conservative approaches lead to higher operation cost. But depending on the chosen cut-point, the proposed approach can sometimes lead to better frequency response quality, while keeping the operation costs relatively low. As it can be seen in table \ref{tresults}, more conservative cut-points lead to less percentage of unacceptable results. Each column in the table is compared with the conventional approach. The ones that perform better than the conventional approach are identified with red boxes, and the ones that perform worse are identified with yellow boxes. The results also show that the proposed approach can guarantee a better frequency response quality if a proper cut-point is chosen. Depending on the required level of cautiousness, the operator can choose a cut-point. For the La Palma island, a probability assurance of $\psi=-6.91$ seems appealing, because both frequency response quality and operation cost are improved.\\
\begin{table*}[tp]
\caption{Results of the simulations on La Palma island}
\label{tresults}
\centering
%\includegraphics[width=\textwidth]{Paper_2_on_IEEE_Journal_Paper_Template (7)_Page_09T.png}
%\begin{comment}
\begin{tabularx}{\linewidth}{g | s s s s s | n | n}
\toprule
& \centering acceptable (\%) & \centering unacceptable (\%) & \centering average $f^{qss}$ & \centering  average $f^{nadir}$ & \centering  average RoCoF & \centering  average UFLS &  \centering operation cost \tabularnewline
\hline
\rowcolor{myTeal}
\centering Conventional approach & \centering$73.1\%$    & \centering $26.9\%$ & \centering 49.61 \scriptsize{Hz} & \centering 48.29 \scriptsize{Hz} & \centering -0.39 \scriptsize{Hz/s}  & \centering 2.30 \scriptsize{MW} & \centering 140.61 \scriptsize{k\euro} \tabularnewline
\hline
\centering LR with $\psi\!=\!2.12$& \tikzmark{top left 1} \centering$83.2\%$  & \tikzmark{top left 2} \centering$16.8 \%$ & \tikzmark{top left 3} \centering 49.81 \scriptsize{Hz} & \tikzmark{top left 4}
\centering 48.81 \scriptsize{Hz} & \tikzmark{top left 5} 
\centering -0.30 \scriptsize{Hz/s} & \tikzmark{top left 6}
\centering 1.14 \scriptsize{MW} (-50.5\%)  & \tikzmark{top left 7}
\centering 145.26 \scriptsize{k\euro} (+3.3\%) \tabularnewline
\rowcolor{myTeal}
\centering LR with $\psi\!=\!0$& \centering$81.4\%$ & \centering$18.6\%$  & \centering 49.82 \scriptsize{Hz}  & \centering 48.77 \scriptsize{Hz} & \centering -0.31 \scriptsize{Hz/s}  & \centering 1.23 \scriptsize{MW}(-46.5\%) & \centering 143.68 \scriptsize{k\euro} (+2.1\%)\tabularnewline
\centering LR with $\psi\!=\!-2.12$& \centering$81.2\%$ & \centering$18.8\%$ & \centering 49.81  \scriptsize{Hz}  & \centering 48.68 \scriptsize{Hz} & \centering -0.33 \scriptsize{Hz/s}  & \centering 1.47 \scriptsize{MW} (-36.1\%) 
&  \centering 142.90 \scriptsize{k\euro}(+2.3\%) \tabularnewline
\rowcolor{myTeal}
\centering LR with $\psi\!=\!-4.95$&\centering$78.2\%$   &\centering$21.8\%$  & \centering 49.74 \scriptsize{Hz}  & \centering 48.57 \scriptsize{Hz} & \centering -0.34 \scriptsize{Hz/s}  &  \centering 1.75 \scriptsize{MW}(-23.9\%) & \centering 141.32 \scriptsize{k\euro} (+0.5\%)\tabularnewline

\centering LR with $\psi\!=\!-5$& \centering$74.9\%$   & \centering$25.1\%$ & \centering 49.73 \scriptsize{Hz}  & \centering 48.48 \scriptsize{Hz} & \centering -0.35 \scriptsize{Hz/s} & \centering 2.03 \scriptsize{MW} (-11.7\%) & \centering 140.78 \scriptsize{k\euro} (+0.1\%)
\tikzmark{bottom right 7}
\tabularnewline

\rowcolor{myTeal}
\centering LR with $\psi\!=\!-6.91$&  \centering$75.1\%$ \tikzmark{bottom right 1} & \centering$24.9\%$ \tikzmark{bottom right 2} &  \centering 49.66  \scriptsize{Hz} \tikzmark{bottom right 3} & \centering 48.52 \scriptsize{Hz}
\tikzmark{bottom right 4}
& \centering -0.35 \scriptsize{Hz/s}  & \centering 2.06 \scriptsize{MW} (-10.4\%)  & \tikzmark{top left 14} \centering 139.83 \scriptsize{k\euro} (-0.6\%) \tabularnewline

\centering LR with $\psi\!=\!-9.21$& \tikzmark{top left 8} \centering$72.4\%$ & \tikzmark{top left 9} \centering$27.6\%$ & \tikzmark{top left 10} \centering 49.26 \scriptsize{Hz}   & \tikzmark{top left 11} \centering 48.09 \scriptsize{Hz}  & \centering -0.37 \scriptsize{Hz/s}
\tikzmark{bottom right 5}
&  \centering 2.20 \scriptsize{MW} (-4.3\%)\tikzmark{bottom right 6} & \centering 138.53 \scriptsize{k\euro} (-1.5\%) \tabularnewline

\rowcolor{myTeal}
\centering LR with $\psi\!=\!-10$& \centering$65.5\%$   & \centering$34.5\%$  &  \centering 47.91 \scriptsize{Hz}  & \centering 46.85 \scriptsize{Hz} & \tikzmark{top left 12} \centering -0.41 \scriptsize{Hz/s}  & \tikzmark{top left 13} \centering 2.44 \scriptsize{MW}(+6.1\%) & \centering 136.86 \scriptsize{k\euro}(-2.7\%) \tabularnewline

\centering LR with $\psi\!=\!-11.51$& \centering$71.4\%$ \tikzmark{bottom right 8}   & \centering$28.6\%$ \tikzmark{bottom right 9} &  \centering 47.97 \scriptsize{Hz}   \tikzmark{bottom right 10} & \centering 46.96 \scriptsize{Hz} \tikzmark{bottom right 11} & \centering -0.38 \scriptsize{Hz/s} \tikzmark{bottom right 12} & \centering 2.61 \scriptsize{MW} (+13.5\%) \tikzmark{bottom right 13} & \centering 136.67 \scriptsize{k\euro} (-2.8\%) \tikzmark{bottom right 14} \tabularnewline
\hline
\rowcolor{myTeal}
\centering OCT, $d$=1, $N$=1001 & \centering$80.4\%$   & \centering$19.6\%$  &  \centering 49.80 \scriptsize{Hz}  & \centering 48.75 \scriptsize{Hz} &  \centering -0.31 \scriptsize{Hz/s}  & \centering 1.31 \scriptsize{MW}(-43.0\%) & \centering 144.33 \scriptsize{k\euro}(+2.6\%) \tabularnewline
\centering OCT, $d$=2, $N$=1001 & \centering$81.8\%$   & \centering$18.2\%$  &  \centering 49.81 \scriptsize{Hz}  & \centering 48.77 \scriptsize{Hz} & \centering -0.31 \scriptsize{Hz/s}  & \centering 1.29 \scriptsize{MW}(-43.9\%) & \centering 145.09 \scriptsize{k\euro}(+3.2\%)\tabularnewline
\rowcolor{myTeal}
\centering OCT, $d$=1, $N$=2800 & \centering  80.9\% & \centering 19.1\% &  \centering 49.80 \scriptsize{Hz}  & \centering 48.76 \scriptsize{Hz} &  \centering -0.31 \scriptsize{Hz/s}  & \centering 1.30 \scriptsize{MW}(-43.5\%) & \centering 144.12 \scriptsize{k\euro}(+2.5\%) \tabularnewline

\bottomrule
\end{tabularx}
%\end{comment}
\end{table*}
%\begin{comment}
\DrawBox[thick, myRed]{top left 1}{bottom right 1}
\DrawBox[thick, myRed]{top left 2}{bottom right 2}
\DrawBox[thick, myRed]{top left 3}{bottom right 3}
\DrawBox[thick, myRed]{top left 4}{bottom right 4}
\DrawBox[thick, myRed]{top left 5}{bottom right 5}
\DrawBox[thick, myRed]{top left 6}{bottom right 6}
\DrawBox[thick, myYellow]{top left 7}{bottom right 7}
\DrawBox[thick, myYellow]{top left 8}{bottom right 8}
\DrawBox[thick, myYellow]{top left 9}{bottom right 9}
\DrawBox[thick, myYellow]{top left 10}{bottom right 10}
\DrawBox[thick, myYellow]{top left 11}{bottom right 11}
\DrawBox[thick, myYellow]{top left 12}{bottom right 12}
\DrawBox[thick, myYellow]{top left 13}{bottom right 13}
\DrawBox[thick, myRed]{top left 14}{bottom right 14}\\
%\end{comment}
To better compare and choose the best $\psi$, all the simulated cases of La Palma island are compared with the conventional approach (highlighted with a yellow cross) in figure \ref{ocufls}.
Although the operation costs go higher by choosing $\psi$ closer to zero, the average UFLS is decreased considerably. Also, there are cases that lead to improvement in both operation cost and average UFLS, which are highlighted in red.\\
The results for OCT in table \ref{tresults}, show improvements in the quality of frequency response compared to the conventional approach and LR with some cut-points. $d$ is the depth of tree structure. OCT with $d=1$ leads to one set of constraints (so the size of UC problem will remain the same), and OCT with $d=2$ leads to two set of constraints. Although OCT is very accurate in classifying the inputs, the run-time of optimization problem relies heavily on the number of inputs and the depth of tree structure. For that reason making the training set smaller was necessary. Solving OCT with full set of training set (around 20000 points) can take many days. So only the biggest hourly outages of some scenarios are considered (like in \cite{lagos2021data}), creating two training data-sets, one smaller with 1001 points, and one bigger with 2800 points. A comparison between the accuracy of representing the data set and solution run-time is presented in table \ref{LRvsOCT}. The down side of a small training set for this practice, is that more unacceptable incidents might be flagged as acceptable and vice versa. As it can be seen in \ref{LRvsOCT}, the advantage of OCT compared to LR is the superior accuracy in classifying the training set and the OCT disadvantage compared to LR is the computational burden of the training process, which effectively limits the size of the training set. Also, tuning the initial values in OCT optimization problem is hard, and affects the ru-time. More discussion about this can be found in \cite{Bertsimas2017}.  For all the simulations in this paper a computer with Intel core i7-8700 CPU, and 32 GB installed RAM is used. \\
\begin{table}[t]
\caption{Comparison of the training process}
\label{LRvsOCT}
\centering
\begin{tabular}{p{0.2\linewidth}|p{0.2\linewidth}p{0.2\linewidth}p{0.2\linewidth}}
\toprule
\centering method & \centering $N$ & \centering error & \centering run-time \tabularnewline
\hline
\rowcolor{myTeal}
\centering LR & \centering 19860 & \centering 3.71\% & \centering 00'03" \tabularnewline
\centering OCT, $d=1$& \centering 1001 & \centering 1.15\% & \centering 00'32" \tabularnewline
\rowcolor{myTeal}
\centering OCT, $d=2$& \centering 1001 & \centering 0.1\% & \centering 28'07" \tabularnewline
\centering OCT, $d=1$& \centering 2800 & \centering 2.07\% & \centering 42'06" \tabularnewline
\bottomrule
\end{tabular}
\end{table}
It's also interesting to see and compare the dynamic frequency responses obtained from the SFR model. In figure \ref{fwithufls} and \ref{fwithoutufls} the frequency response for a period of 15 seconds after outages are presented, for every single outage of online units in a random hour. In figure \ref{fwithufls} the UFLS scheme is activated, and figure \ref{fwithoutufls} shows frequency response with no UFLS. The simulations for the conventional approach are in yellow, the most conservative case with $\psi=2.12$ in red, and one of the preferred cases with $\psi=-6.91$ in green. The moments that the UFLS scheme has operated are also highlighted with dashes. The better performance of the conservative case is noticeable. Also, the case with $\psi=-6.91$ outperforms the conventional approach. The minimum allowed frequency nadir is shown with the gray line in figure \ref{fwithoutufls}.
\section{Conclusion}\label{conclution}
This paper proposes a novel procedure to schedule short-term unit commitment in island power systems. Island power systems usually suffer from lack of inertia and frequency response capacity, complicating containing frequency within an acceptable range during large disturbances. The proposed method uses an initial data set to train a linear constraint that takes into account the dynamic response of the system. For the purpose of training this constraint, logistic regression is employed to avoid incidents with undesirable frequency responses as much as possible. Then the logistic regression constraint is included in an adaptive robust formulation. Results show that by choosing a proper cut-point, the proposed method improves the frequency response, as well as the operation costs. As training data with LR model is very fast, the size of training set is not an issue. A complete training data set can better represent the system, leading to a more reliable frequency constraint.

\ifCLASSOPTIONcaptionsoff
  \newpage
\fi

\bibliographystyle{IEEEtran}
\bibliography{paper2IEEE}

%\begin{IEEEbiography}{Mohammad Rajabdorri}
%\end{IEEEbiography}

%\begin{IEEEbiographynophoto}{Lukas Sigrist}

%\newpage

%\begin{IEEEbiographynophoto}{Enrique Lobato}


%\vfill

\end{document}